# COALESCING AT 8 GEV IN THE FERMILAB MAIN INJECTOR

D. J. Scott, D. Capista, B. Chase, J. Dye, I. Kourbanis, K. Seiya, M.-J. Yang, Fermilab, Batavia, IL 60510, U.S.A.

*Abstract*

For Project X, it is planned to inject a beam of $3 \times 10^{11}$ particles per bunch into the Main Injector. To prepare for this by studying the effects of higher intensity bunches in the Main Injector it is necessary to perform coalescing at 8 GeV. The results of a series of experiments and simulations of 8 GeV coalescing are presented. To increase the coalescing efficiency adiabatic reduction of the 53 MHz RF is required. This results in ~70% coalescing efficiency of 5 initial bunches. Data using wall current monitors has been taken to compare previous work and new simulations for 53 MHz RF reduction, bunch rotations and coalescing, good agreement between experiment and simulation was found. By increasing the number of bunches to 7 and compressing the bunch energy spread a scheme generating approximately $3 \times 10^{11}$ particles in a bunch has been achieved. These bunches will then be used in further investigations.

## INTRODUCTION

Coalescing is a non-adiabatic process that can be used to increase bunch intensities [1]. In the Fermilab Main Injector (MI) five 53 MHz bunches are rotated, via synchrotron oscillations, in a 2.5 MHz bucket and then recaptured in a 53 MHz RF bucket. The basic manipulations are shown in Figure 1 and Table 1 gives some basic parameters for standard operations and those expected for Project X.

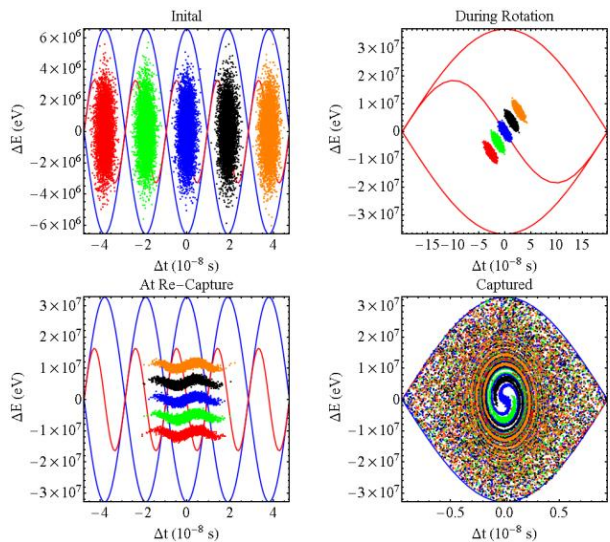

Figure 1: Coalescing schematic. 5 bunches initially in 53 MHz Buckets (blue lines) rotate in a 2.5 MHz bucket (red line) then are recaptured in the 53 MHz RF, the particles then mix in the central 53 MHz bucket.

The coalescing efficiency, $\alpha$, is the ratio of captured to initial particles. This strongly depends on the energy spread, $\sigma_E$, of the beam before rotation, the bunch lengths, $\sigma_t$, have little to no influence. Figure 2 shows the results of simulations calculating $\alpha$ for different initial $\sigma_E$ and $\sigma_t$ and 5 bunches The typical beam injected into the MI has $\sigma_E = 3$ MeV giving $\alpha = 63$ %. Previous studies [2] attempted to achieve $\alpha = 85$ % by reducing the energy spread using bunch compression with the 53 MHz RF and bunch stretching on the unstable fixed point during rotation in the 2.5 MHz RF. To implement bunch stretching with the MI requires modifications to the timing resolution of the low level RF (LLRF) system are required, that have yet to be implemented.

Table 1: Beam Power in the MI

|  | Unit | Operations | Project X |
|---|---|---|---|
| Beam Power | kW | 400 | 2000 |
| Total Intensity | $10^{14}$ | 0.4 | 1.6 |
| # of Bunches |  | 492 | 458 |
| Bunch Intensity | $10^{11}$ | 1.0 | 3.0 |
| MI Cycle Time | s | 2.2 | 1.4 |

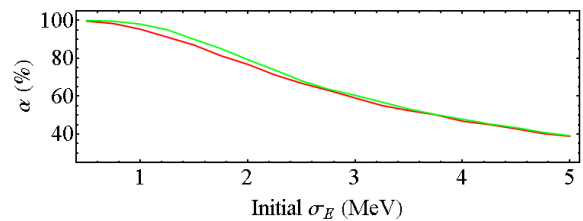

Figure 2: $\alpha$ vs $\sigma_E$ with $\sigma_t = 1$ and 9 ns (red, green).

This paper will outline the results of experiments and simulations that have achieved coalesced bunch intensities of $3 \times 10^{11}$ particles using the existing MI LLRF. This has been achieved through using 7 initial bunches of the highest intensity available from the Fermilab Booster and allowing the beam to rotate in the 2.5 MHz RF for 0.75 rotations, instead of 0.25, to allow for the current minimum time between LLRF commands.

## DETERMINING THE BEAM ENERGY SPREAD

To compare experimental results with simulations an estimation of $\sigma_E$ is required. This has been achieved by using Wall Current Monitor (WCM) measurements of the beam intensity vs longitudinal position around the ring during compression of $\sigma_E$. Compressing $\sigma_E$ is achieved by adiabatically reducing the 53 MHz voltage from the nominal value of 1.1 MV. During this manipulation $\sigma_t$ increases and that can be measured from the WCM data. Figure 3 shows an example of the bunch length increasing as the 53 MHz RF is adiabatically reduced from 1.1 MV to 45 kV over 0.2 seconds. The measured values of $\sigma_t$ can be compared to simulations for a given $\sigma_E$, a good match indicating a reasonable estimation of $\sigma_E$. The simulation and experimental results are shown in Figure 4 using $\sigma_E$ values given in Table 2. The simulation deviates from the measurements near the end of the compression for the lower intensity bunch data. This could be due to the WCM data becoming more noisy as the peak signal decreases and the fits underestimate the bunch length. The longitudinal beam emittance, $\varepsilon$, can be estimated using $\varepsilon = 4\pi\sigma_t\sigma_E$ and this is also given in Table 2. The expected emittance from the booster is 0.1 eV s indicating that the low intensity bunches are not well matched, which is expected as nominal operations uses the higher, $5\ 10^{11}$, bunch intensities.

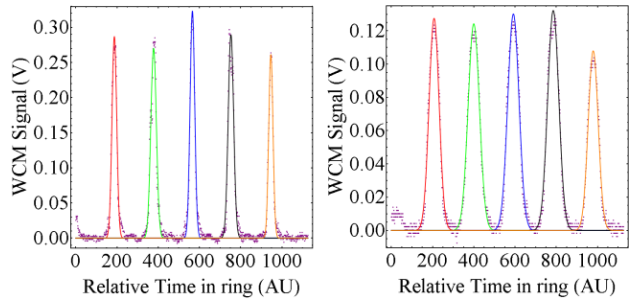

Figure 3: WCM data, points, for initial bunches (left) and after $\sigma_E$ compression (right). Gaussian fits for each bunch are also shown, used to determine $\sigma_t$.

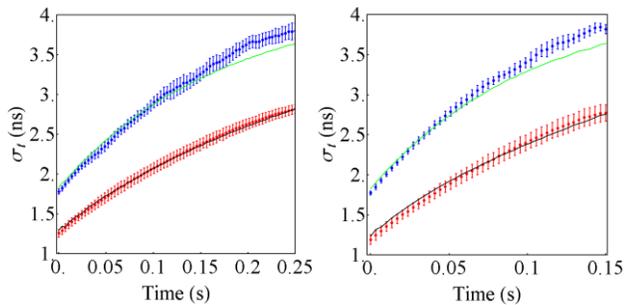

Figure 4: Mean $\sigma_t$ for 0.25 and 0.15 s compression for $5\ 10^{10}$ (red, black) and $1\ 10^{10}$ (blue, green) bunch intensities.

Table 2: Longitudinal emittance and $\sigma_E$ determined by simulations.

| Bunch Intensity | Initial $\sigma_E$ (MeV) | Final $\sigma_E$ (MeV) | $\varepsilon$ (eV s) |
|---|---|---|---|
| $1\ 10^{10}$ | 9.5 | 3.3 | 0.22 |
| $5\ 10^{10}$ | 6.7 | 2.8 | 0.11 |

## ROTATION IN THE 2.5 MHZ RF

The rotation time in the 2.5 MHz bucket depends on the voltage, and can be calculated from the synchrotron tune. This has been measured and compared with theory. Figure 5 shows a typical contour plot of WCM data for 5 rotating bunches. The rotation time can be found by plotting the peak signal for each turn of WCM data and then finding the time between maximums of these peaks, as shown in Figure 6. Figure 6 also shows the simulated and measured rotation times for different values of the 2.5 MHz RF voltage. There is excellent agreement between the two.

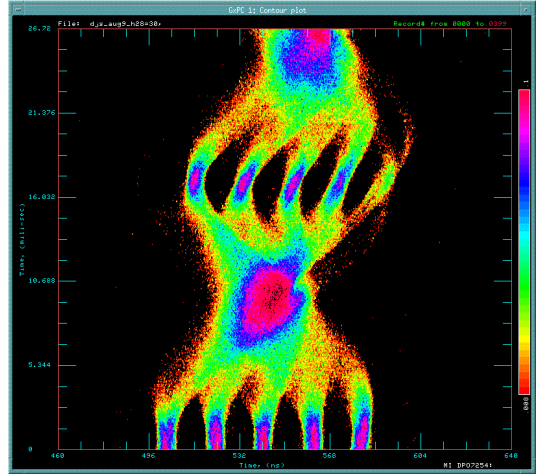

Figure 5: Contour plot of the WCM signal data showing 5 initial bunches rotating in the 2.5 MHz RF. The horizontal axis is longitudinal position in the MI.

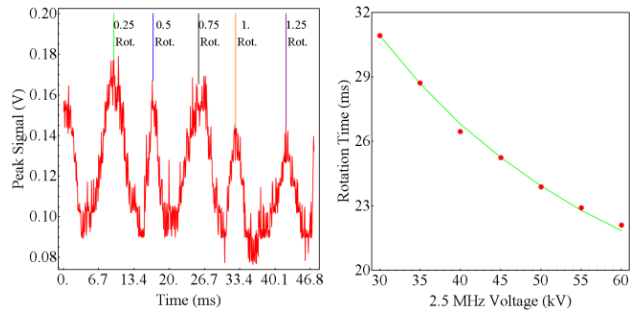

Figure 6: (Left) example calculation of rotation time in 2.5 MHz bucket showing the peak of the WCM signal for each turn of data. (Right) simulated (line) and measured (points) rotation time for different 2.5 MHz RF voltages.

## COALESCING WITH EXISTING RF

The timing limitations of the MI LLRF were overcome by para-phasing the 53 MHz RF whilst the 2.5 MHz was on in order to set the effective voltage to zero. The bunches were allowed to rotate for 0.75 rotations or synchrotron periods, (instead of the minimum 0.25) in the 2.5 MHz RF in order to have enough time between LLRF commands. After 0.75 rotations the 2.5 MHz is switched off and the 53 MHz is snapped back on, again with para-phasing. Figure 7 shows the 53 and 2.5 MHz RF voltages and the beam current in the machine for this coalescing scheme. Figure 8 shows an example WCM contour plot of successful coalescing. Here 7 initial bunches start rotating for about 15 ms in the 2.5 MHz RF (0.75 rotations). They are then recaptured in the 53 MHz as one high intensity middle bunch with two smaller intensity satellites either side.

### Profile at Recapture

There is a good agreement between the simulated and measured bunch profiles at recapture, shown in Figure 9.

Expected increases in intensity due to increased bunch compression can also been seen.

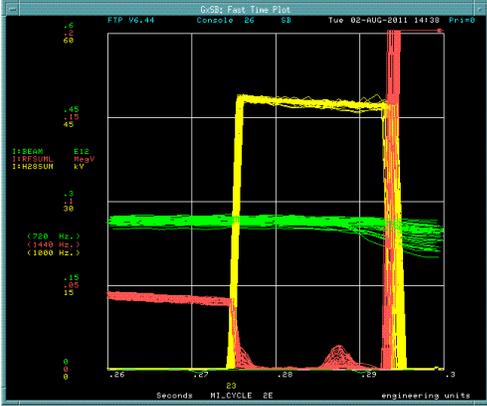

Figure 7: 53 MHz voltage, (red), 2.5 MHz RF voltage (yellow) and Beam current (green) during coalescing.

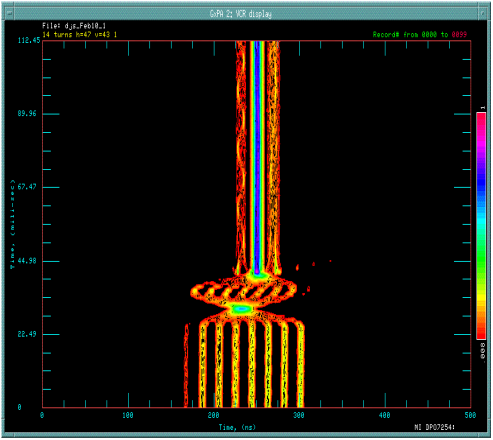

Figure 8: WCM contour plot showing an example of coalescing with 7 initial bunches.

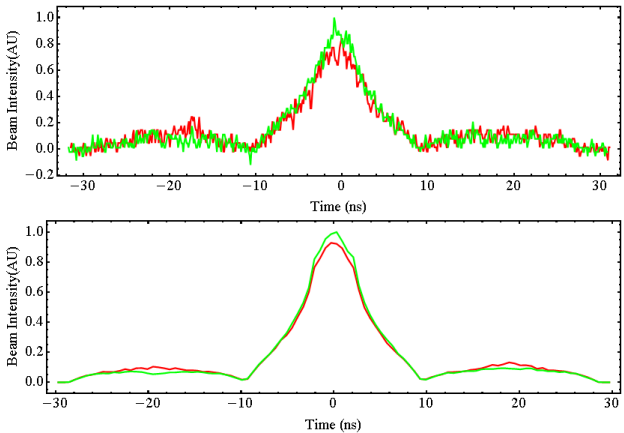

Figure 9: Beam profile for WCM data (top) and simulations (bottom) when 53 MHz is reduced to 30 (red) and 25 (green) kV.

*Coalescing Efficiency*

Figure 10 shows a comparison between simulations and measurements of the expected coalescing efficiency for 5 and 7 bunches and different 53 MHz RF voltages after compression, i.e. different $\sigma_E$ values. The reduction in efficiency as the 53 MHz is reduced below 15 kV is expected as the beam emittance becomes comparable to the bucket area at around this voltage and so particles are lost.

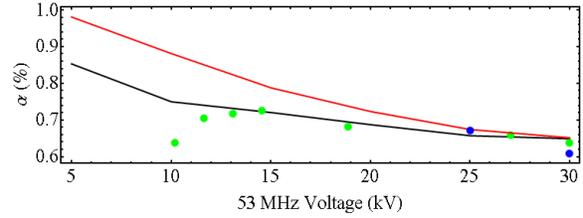

Figure 10: Simulated, (lines) and measured (points) coalescing efficiency for 5 (red, blue) and 7 (green black) bunches

*Number of Particles in Coalesced Bunch*

The number of particles in the bunch was found by normalising the integrated WCM signal with beam current measurements from the MI control system. The number of particles in the central coalesced bunches are shown in Figure 11 for over 300 different events.

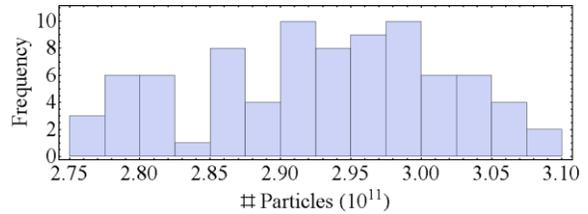

Figure 11: Coalesced bunch intensities.

## CONCLUSION

It has been shown that bunches with $3 \cdot 10^{11}$ particles can be generated in the MI using coalescing at 8 GeV and the existing RF. Throughout, the experiments have well matched simulations. There is a spread of bunch intensities of ± 5 % between different measurements. This is attributed to effects that are difficult to control, such as alignment of the two RF systems drifting over time. As this experiment is in an initial phase there are, as yet, no good online diagnostic and analysis tools to help maintain a consistent coalescing efficiency. These high intensity bunches have been used to study space charge tune shifts in the MI in preparation for Project X [3].